\documentclass{article}
 
    \usepackage[preprint]{neurips_2023}

\usepackage[utf8]{inputenc} 
\usepackage[T1]{fontenc}    
\usepackage{hyperref}       
\usepackage{url}            
\usepackage{booktabs}       
\usepackage{amsfonts}       
\usepackage{nicefrac}       
\usepackage{microtype}      
\usepackage{xcolor}         
\usepackage{graphicx}
\usepackage{tabularx}  
\usepackage{booktabs}  
\usepackage{array}     
\usepackage{authblk}
\usepackage{amsmath}
\usepackage{booktabs}
\usepackage{enumitem}
\usepackage{amsfonts}
\usepackage{amsmath}

\usepackage{multirow}

\usepackage{algorithmic}
\usepackage[linesnumbered,ruled,vlined]{algorithm2e}

\title{SqueezeComposer: Temporal Speed-up is A Simple Trick for Long-form Music Composing}
\makeatletter
\renewcommand\AB@affilsepx{ \protect\Affilfont} 
\renewcommand\AB@affilnote[1]{\mbox{\textsuperscript{#1}}}
\makeatother

\author{
  \textbf{Jianyi Chen\textsuperscript{1}},
  \textbf{Rongxiu Zhong\textsuperscript{2}},
  \textbf{Shilei Zhang\textsuperscript{2}},
  \textbf{Kun Qian\textsuperscript{3}},
  \textbf{Jinglei Liu\textsuperscript{4}},
  \textbf{Yike Guo\textsuperscript{1}},
  \textbf{Wei Xue\textsuperscript{1\dag}} 
\\
  \textsuperscript{1}The Hong Kong University of Science and Technology, \\
  \textsuperscript{2}JIUTIAN Research of China Mobile, \\
  \textsuperscript{3}Beijing Institute of Technology, \\
  \textsuperscript{4}China Mobile (Hong Kong) Innovation Research Institute
}

\begin{document}

\maketitle

\begin{abstract}
Composing coherent long-form music remains a significant challenge due to the complexity of modeling long-range dependencies and the prohibitive memory and computational requirements associated with lengthy audio representations. In this work, we propose a simple yet powerful trick: we assume that AI models can understand and generate time-accelerated (speeded-up) audio at rates such as 2$\times$, 4$\times$, or even 8$\times$. By first generating a high-speed version of the music, we greatly reduce the temporal length and resource requirements, making it feasible to handle long-form music that would otherwise exceed memory or computational limits. The generated audio is then restored to its original speed, recovering the full temporal structure. This temporal speed-up and slow-down strategy naturally follows the principle of hierarchical generation from abstract to detailed content, and can be conveniently applied to existing music generation models to enable long-form music generation. We instantiate this idea in SqueezeComposer, a framework that employs diffusion models for generation in the accelerated domain and refinement in the restored domain. We validate the effectiveness of this approach on two tasks: long-form music generation, which evaluates temporal-wise control (including continuation, completion, and generation from scratch), and whole-song singing accompaniment generation, which evaluates track-wise control. Experimental results demonstrate that our simple temporal speed-up trick enables efficient, scalable, and high-quality long-form music generation. Audio samples are available at \url{https://SqueezeComposer.github.io/}.

\end{abstract}


\section{Introduction}

Recent advances in audio generation have enabled the creation of high-quality music using autoregressive language models~\cite{agostinelli2023musiclm, copet2023simple, donahue2023singsong, chen2024pyramidcodec, lam2023efficient} and diffusion models~\cite{copet2023simple, schneider2024mousai, huang2023noise2music, dhariwal2020jukebox, zhu2023ernie, evans2024long, chen2024musicldm}. While these models demonstrate impressive performance in generating short musical segments with coherent local structure and high audio fidelity, extending them to long-form music—spanning several minutes—remains a significant challenge. This difficulty is two-fold. First, the computational and memory requirements grow rapidly with representation, making it infeasible to directly model or generate extremely long audio using current architectures. Second, maintaining global musical coherence and effectively capturing long-range dependencies over extended durations is inherently difficult, often resulting in outputs that lack structural consistency or thematic continuity. To address these challenges, learning more compact audio representations and adopting a hierarchical composition strategy offer promising directions for scalable long-form music generation.

\begin{figure}[!t]
\centering
\includegraphics[width=0.8\columnwidth]{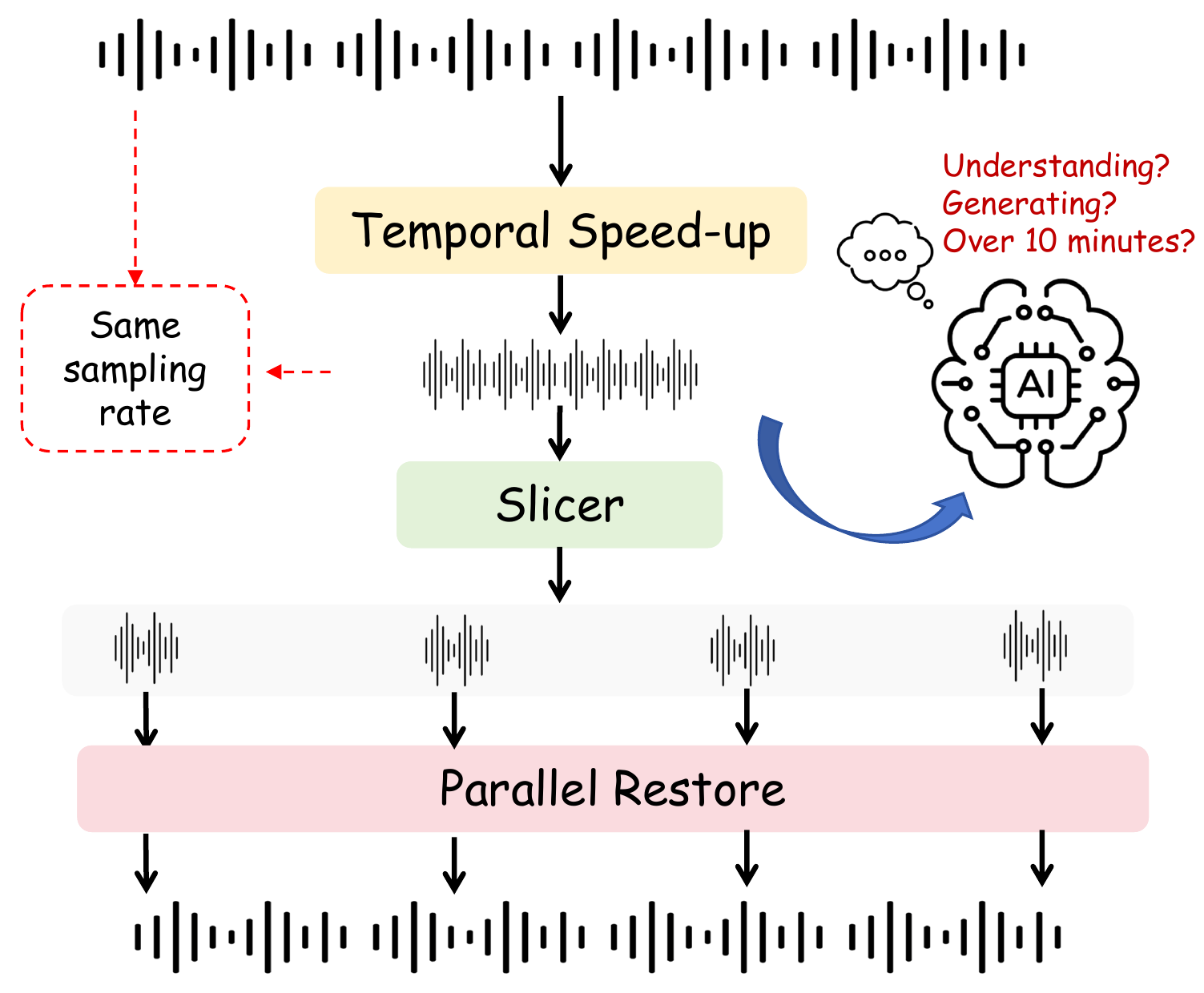}
\caption{The SqueezeComposer illustration: audio is compressed via time-scale modification, generated in a compact domain for various tasks (e.g., long-form composition, accompaniment), and then expanded and refined to the original resolution.}
\label{fig:overview}
\end{figure}


Since directly generating raw long-form audio waveforms is computationally prohibitive, mainstream approaches tend to use intermediate representations to bridge between the audio and generative models, which can be continuous or discrete. Continuous VAE-derived latent features are adopted in the Latent diffusion models~\cite{evans2024long, chen2024musicldm, evans2024fast, liu2024audioldm, xu2024multi}, and Mel-spectrogram representations are also taken as the generation target in other diffusion-based frameworks \cite{copet2023simple, schneider2024mousai, huang2023noise2music}. Discrete audio codecs are used in language model-based frameworks \cite{agostinelli2023musiclm, copet2023simple, donahue2023singsong, chen2024pyramidcodec, lam2023efficient, yang2024uniaudio} for autoregressive or masked generation. For all approaches, we note that the representations are extracted at the original audio speed. However, since the audio contains much low-level information, a much higher temporal modeling resolution is required compared to text language modeling. For instance, for the commonly adopted 10~ms hop size for 80-bin Mel-spectrogram modelling, a 5-minute audio corresponds to a representation size as 30,000$\times$80, making the direct diffusion modelling difficult. Generally, the token-per-second (TPS) rate for audio tokenizers ranges from 25 to 200, thus even considering only a single codebook, at least 25$\times$300=7,500 tokens are needed for the sequential modeling. Therefore, because the sequence length is proportional to the temporal length, traditional frameworks lead to excessive redundancy and pose challenges for efficient long-term generation and structural modeling.

Several recent studies have sought to reduce the length of audio representations to enable more efficient modeling. For continuous features, Mustango~\cite{melechovsky2024mustango} learns VAE-based latent representations directly from Mel spectrograms instead of raw waveforms, achieving a more compact and manageable representation space. Similarly, ACE-Step~\cite{gong2025ace} employs a Deep Compression AutoEncoder (DCAE)~\cite{chen2024deep} to obtain 8 times denser Mel-spectrogram features, integrating a diffusion model within a complex network architecture to further improve reconstruction fidelity. For discrete features, UniCodec~\cite{jiang2025unicodec} introduces a unified multi-domain codec that operates at 75 TPS using a Mixture-of-Experts strategy across domains, while MuCodec~\cite{xu2024mucodec} extracts ultra-low-rate tokens (25 TPS) through a diffusion-based codec to preserve high audio quality. PyramidCodec~\cite{chen2024pyramidcodec} and SNAC~\cite{siuzdak2024snac} further adopts a hierarchical design, operating at only 10 TPS at its highest abstraction level, though the total sequence length increases when all hierarchical layers are utilized. Overall, these approaches effectively shorten sequence length while maintaining perceptual fidelity, but often depend on specialized architectures, complex training schemes, or retraining on large-scale datasets. 

Despite many studies aiming to propose the low-rate representations, here we explore another perspective for effective audio modelling. We note that unlike the text which cannot be directly ``squeezed'' over time, the ``squeezed'', i.e., accelerated audio at certain rates, can be still understood. Therefore, we propose a simple yet effective perspective (the basic idea is illustrated in Figure~\ref{fig:overview}): we hypothesize that AI models can understand and generate \textit{accelerated} audio over time at rates such as $4\times$, $8\times$, or even $16\times$ faster than real time. Instead of using original-speed audio, the accelerated audios are used to generate intermediate representations, which are further utilized to train the generation models. Even the operation is simple, this would immediately reduce the sequence length and computational requirements, making it feasible to model long-form music which can even last several dozens of minutes. The generated audio is then restored to its original playback speed, recovering the full temporal structure. This {acceleration and restoration} strategy is intuitive, model-agnostic, and aligns naturally with the principle of hierarchical generation: the accelerated domain represents the abstract, coarse-grained musical structure, while the restoration stage refines it into detailed, high-fidelity audio. 
We emphasize that our squeeze-and-restore operation is applied at the mel-spectrogram level.
Although existing vocoders are typically trained on mel-spectrograms derived from natural-speed audio, we find that mel representations obtained after squeeze and restore remain compatible with off-the-shelf vocoders(e.g.,~\cite{kong2020hifi, lee2023bigvgan}) without requiring retraining. While the use of restored mel-spectrograms may introduce some degradation compared to ground-truth mel inputs, the resulting audio quality remains within an acceptable range in our experiments.


The above discussion leads to the proposed {SqueezeComposer}, a hierarchical framework that generates music in the accelerated domain and refines it by restoring to the original temporal scale. To ensure the overall efficiency of the framework in music generation, here, we use diffusion models throughout the process to demonstrate the effectiveness of squeezing audios. We evaluate SqueezeComposer on two representative tasks: long-form music generation, which evaluates temporal-wise generation (including continuation, completion, and generation from scratch), and whole-song singing accompaniment generation, which evaluates track-wise generation. Experimental results demonstrate that our temporal speed-up trick enables efficient and high-quality long-form music generation. 

Our main contributions are summarized as follows:
\begin{itemize}
    \item We propose \textit{SqueezeComposer}, a general and model-agnostic framework for long-form music generation that leverages audio acceleration to reduce sequence length and computational cost while preserving musical coherence.
    \item We formulate a hierarchical generation paradigm, where the accelerated domain serves as an abstract representation capturing global musical structure, and the restoration process refines it into detailed, high-fidelity audio.
    \item We design diffusion-based methods in the Mel-based continuous domain and validate them on two representative tasks: (1) long-form music generation (temporal-wise control) and (2) whole-song singing accompaniment generation (track-wise control), demonstrating effective temporal and harmonic modeling.
\end{itemize}

\section{Related Work}

\subsection{Audio Representations}
Audio representations form the foundation of music generation systems, bridging the gap between raw waveforms and high-level modeling. Traditional vocoders~\cite{kumar2019melgan, kong2020hifi, lee2023bigvgan, siuzdak2023vocos} reconstruct audio from continuous spectral features such as Mel spectrograms, offering high-fidelity synthesis and efficient conditioning. To further enhance compactness, several works propose using VAE-based latent representations as intermediate features~\cite{gong2025ace, chen2024musicldm, liu2024audioldm, evans2024long}, enabling generative modeling in compressed continuous spaces.  Discrete representations, in contrast, quantize audio into token sequences through learned codecs~\cite{zeghidour2022soundstream, defossez2023high, kumar2023high, xu2024mucodec, pepino2025encodecmae, wu2023audiodec, li2025melcap, wang2025switchcodec, zhang2025mbcodec, wu2025ts3, zhai2025one, jiang2025unicodec, xu2024mucodec, chen2024pyramidcodec}. These codecs typically operate at tens or hundreds of TPS and vary in structure.  While these continuous and discrete representations greatly improve scalability and audio quality, they are all extracted from audio at its original playback speed. Consequently, the sequence length of their features grows linearly with music duration, leading to long, flat sequences that are difficult to model for minute-scale generation. This inherent limitation motivates the development of our time-accelerated framework, which directly shortens the temporal length of audio representations before modeling.


\subsection{Music Generation}

Music generation has been studied across multiple representational domains, most notably symbolic music and audio. Early and ongoing work in symbolic-domain music generation focuses on structured representations such as notes, chords, and scores~\cite{roberts2018hierarchical, dong2018musegan, huang2019music, yu2022museformer, hsiao2021compound, mittal2021symbolic}. In parallel, a growing body of research explores music generation directly in the audio domain, which enables richer acoustic modeling at the cost of significantly higher temporal resolution and computational complexity. Transformer-based autoregressive methods~\cite{agostinelli2023musiclm, copet2023simple, donahue2023singsong, dhariwal2020jukebox, lam2023efficient} exhibit strong local coherence, but in practice they can only handle limited music duration because the quadratic attention cost and finite computation resources constrain the context window. Other autoregressive methods adopt state-space models (SSMs)~\cite{gu2022efficiently, gu2020hippo, gu2024mamba} for audio or music generation~\cite{goel2022s, lee2025exploring, yuan2025diffusion, erol2024audio}, reducing the quadratic complexity of attention to linear while maintaining the long-range modeling capability. Diffusion-based generative methods~\cite{liu2024audioldm, chen2024musicldm, evans2024long, xu2024multi} are also developed to iteratively synthesize audio in continuous latent space or Mel-spectrogram. To improve long-range coherence, hierarchical and multi-stage frameworks have been proposed: Jukebox~\cite{dhariwal2020jukebox} adopts a multi-level VQ-VAE~\cite{van2017neural} to capture coarse-to-fine musical structure, while AudioLM~\cite{borsos2023audiolm} and MusicLM~\cite{agostinelli2023musiclm} combine neural codecs with language modeling to separate semantic, acoustic, and fine-grained levels for more coherent long-form synthesis. Recent works have also focused on performing information compression in learned codecs~\cite{jiang2025unicodec, xu2024mucodec, gong2025ace, chen2024pyramidcodec} to reduce sequence length; however, they still rely on dedicated algorithm design and operate at full-speed audio. In contrast, our framework introduces a complementary perspective: by accelerating the audio before feature extraction, we can straightforwardly achieve implicit temporal compression of both continuous and discrete representations, enabling the generation of significantly longer and more coherent musical pieces within practical computational budgets.

\subsection{Singing Accompaniment Generation}
Besides whole music generation, singing accompaniment generation aims to automatically produce instrumental tracks conditioned on a given vocal melody, with applications in music production and karaoke. However, generating a complete song often takes several minutes and requires precise rhythmic synchronization, harmonic consistency, and strong long-term modeling. Early works such as SingSong~\cite{donahue2023singsong} used transformer-based autoregressive models for short vocal-conditioned accompaniments, while FastSAG~\cite{chen2024fastsag} employed diffusion-based parallel generation for faster inference. Despite these advances, most existing methods are limited to short clips (about 10 seconds) and fail to maintain coherence across full-length songs, primarily due to the computational burden of modeling minute-scale multi-track structures. To address this challenge, our audio acceleration based framework directly facilitates scalable, whole-song accompaniment generation with consistent rhythmic and harmonic structure.

\begin{algorithm}
\caption{SqueezeComposer}
\label{alg:SqueezeComposer}
\begin{algorithmic}[1]
\REQUIRE Input audio $x$, speed-up ratio $r$, feature extractor $\mathcal{E}$, generator $\mathcal{G}$, expansion model $\mathcal{G}_{\text{exp}}$, vocoder $\mathcal{D}$
\STATE $x_c \leftarrow \text{TimeScaleCompress}(x, r)$
\STATE $z_c \leftarrow \mathcal{E}(x_c)$
\STATE $c \leftarrow \text{PrepareConditioning}(z_c, \text{task})$
\STATE $\hat{z}_c \leftarrow \mathcal{G}(c)$ \COMMENT{High-level generation in compressed domain}
\STATE $\hat{z} \leftarrow \text{TimeScaleExpand}(\hat{z}_c, r)$
\STATE $\tilde{z} \leftarrow \mathcal{G}_{\text{exp}}(\hat{z})$ \COMMENT{Detail refinement}
\STATE $\hat{x} \leftarrow \mathcal{D}(\tilde{z})$
\RETURN $\hat{x}$
\end{algorithmic}
\end{algorithm}

\section{Squeezing and Restoring}

\subsection{Overview}
SqueezeComposer leverages a simple temporal speed-up trick for efficient and scalable long-form music generation. The core idea is to first accelerate the audio signal (e.g., 2$\times$, 4$\times$, or 8$\times$ faster), extract intermediate representations from this accelerated audio, and then generate music in this compact domain before restoring it to the original speed. This compress-and-expand paradigm addresses the challenges of long-form generation by significantly reducing memory and computational requirements, enabling fast generation with non-autoregressive diffusion models, and effectively modeling long-range dependencies via hierarchical, abstract-to-detail generation. Instead of extracting intermediate representations directly from the original audio—which results in long and computationally expensive features—SqueezeComposer first compresses the audio using time-scale modification, performs high-level generation in this compact domain, and then expands the generated representation back to the original temporal resolution with additional detail refinement. The basic paradigm of SqueezeComposer is shown in Algorithm~\ref{alg:SqueezeComposer}. This process enables both temporal-wise and track-wise scalable music generation.

\subsection{Preliminary}
In this section, we introduce the score-based diffusion model~\cite{song2021score}, which is widely used in audio or music generation framework such as~\cite{liu2024audioldm, evans2024long} and serves as our conditional probabilistic framework for both the temporal restoration of accelerated audio and the continuous-domain music generation.

Let the data distribution be $p_{\text{data}}(\mathbf{x})$, and define a family of distributions $p(\mathbf{x}; \delta)$ obtained by adding Gaussian noise $\mathcal{N}(0, \delta I)$ to the data samples. The VE diffusion process begins by sampling a noisy input $\mathbf{x}_0 \sim \mathcal{N}(0, \delta_{\max} I)$ and then sequentially denoising it through a series of decreasing noise levels $\delta_{0} = \delta_{\max} > \delta_{1} > \cdots > \delta_{N} = 0$. After $N$ denoising steps, the final sample $\mathbf{x}_N$ follows a distribution that approximates $p_{\text{data}}(\mathbf{x})$.

Let $D(\mathbf{x}; \delta)$ denote the denoising function that minimizes the expected $\ell_2$ error for each noise level $\delta$, defined as:
\begin{align}
    \mathbb{E}_{\mathbf{y} \sim p_{\text{data}}}\,\mathbb{E}_{\mathbf{n} \sim \mathcal{N}(0, \delta^2 I)} 
    \big\| D(\mathbf{y+n}; \delta) - \mathbf{y} \big\|_2^2.
\end{align}

In our implementation, the denoiser $D(\mathbf{x}; \delta(t))$ is instantiated as a Diffusion Transformer (DiT)~\cite{peebles2023scalable}, expressed as $D_{\theta}(\mathbf{x}_t, \delta(t), \mathit{c})$ in the conditional setting, where $\mathit{c}$ denotes the conditioning input (e.g., features or control signals). This diffusion structure is employed in two parts of SqueezeComposer: 1) Generation: the backbone to compose a series of continuous-domain music representations at an accelerated speed. 2) Restoration: reconstructs the original-speed waveform from accelerated representations generated in the previous step.

\subsection{Temporal Squeezing and Restoring}
\label{restoring}
 Given an input audio waveform, we first apply time-scale compression by accelerating the audio signal with a factor $r$ (e.g., $r=4$). From the accelerated audio, we extract a compact intermediate representation, such as a Mel spectrogram or neural codec features. This representation is significantly shorter compared to features from the original audio, which greatly reduces computational and memory requirements for subsequent modeling.

A conditional diffusion model, which is implemented as DiT here, is then used to generate or complete the music in this compressed domain. After high-level generation, we perform time-scale expansion by restoring the generated representation to its original temporal length, followed by a second diffusion model for detail refinement to enhance the reconstruction quality. This refinement process can be implemented in parallel to accelerate the speed. The final output is decoded into waveform audio using a pretrained vocoder such as BigVGAN.


\begin{figure}[!t]
\centering
\includegraphics[width=0.8\columnwidth]{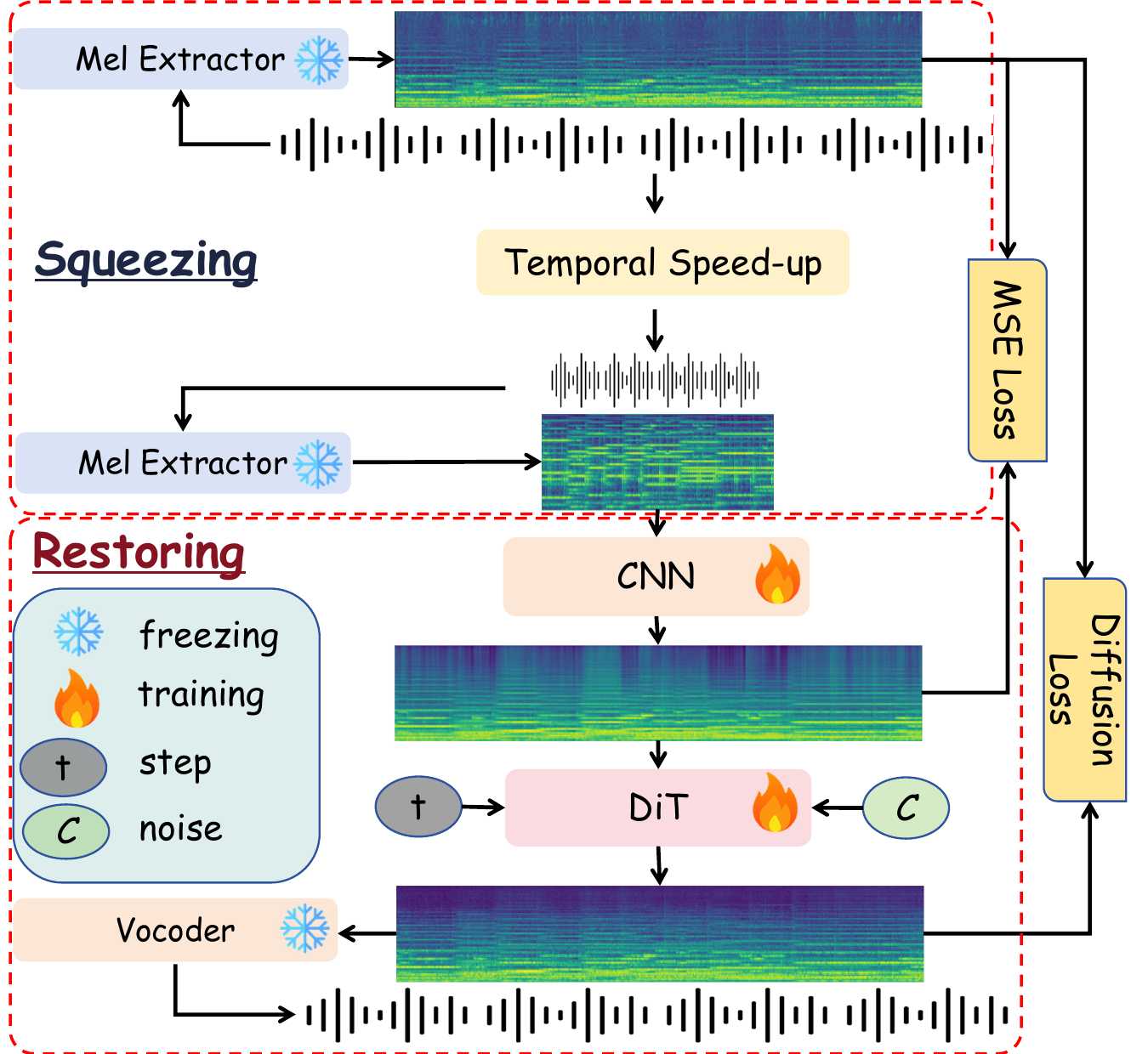}
\caption{Training pipeline for temporal speeding-up and restoration. The input audio is speeded up and processed through a two-stage pipeline: CNN generates a prior condition, then DiT refines it to produce high-quality restored audio. The accelerated audio maintains the original sampling rate, ensuring vocoder compatibility. Training uses MSE loss for CNN prior generation and diffusion loss for DiT refinement.}
\label{fig:train}
\end{figure}

The training pipeline for temporal speeding-up and restoration is illustrated in Figure~\ref{fig:train}. The training process involves learning the temporal speed-up and restoration capabilities. We use the full-length Mel spectrogram from the original audio as ground truth for loss computation. The training pipeline processes the input audio through temporal speed-up (which is a deterministic process and does not require training), followed by a two-stage restoration. In the first stage, a lightweight CNN predicts a coarse prior representation at the original temporal resolution. This CNN serves to upsample and align the temporally compressed Mel representation produced by the speed-up operation to the resolution required by the subsequent diffusion model. In the second stage, a diffusion transformer (DiT) refines this coarse prior to generate a high-quality restored Mel spectrogram. The accelerated audio maintains the same sampling rate as the original, ensuring compatibility with existing vocoders.

\begin{figure*}[!t]
\centering
\includegraphics[width=0.95\textwidth]{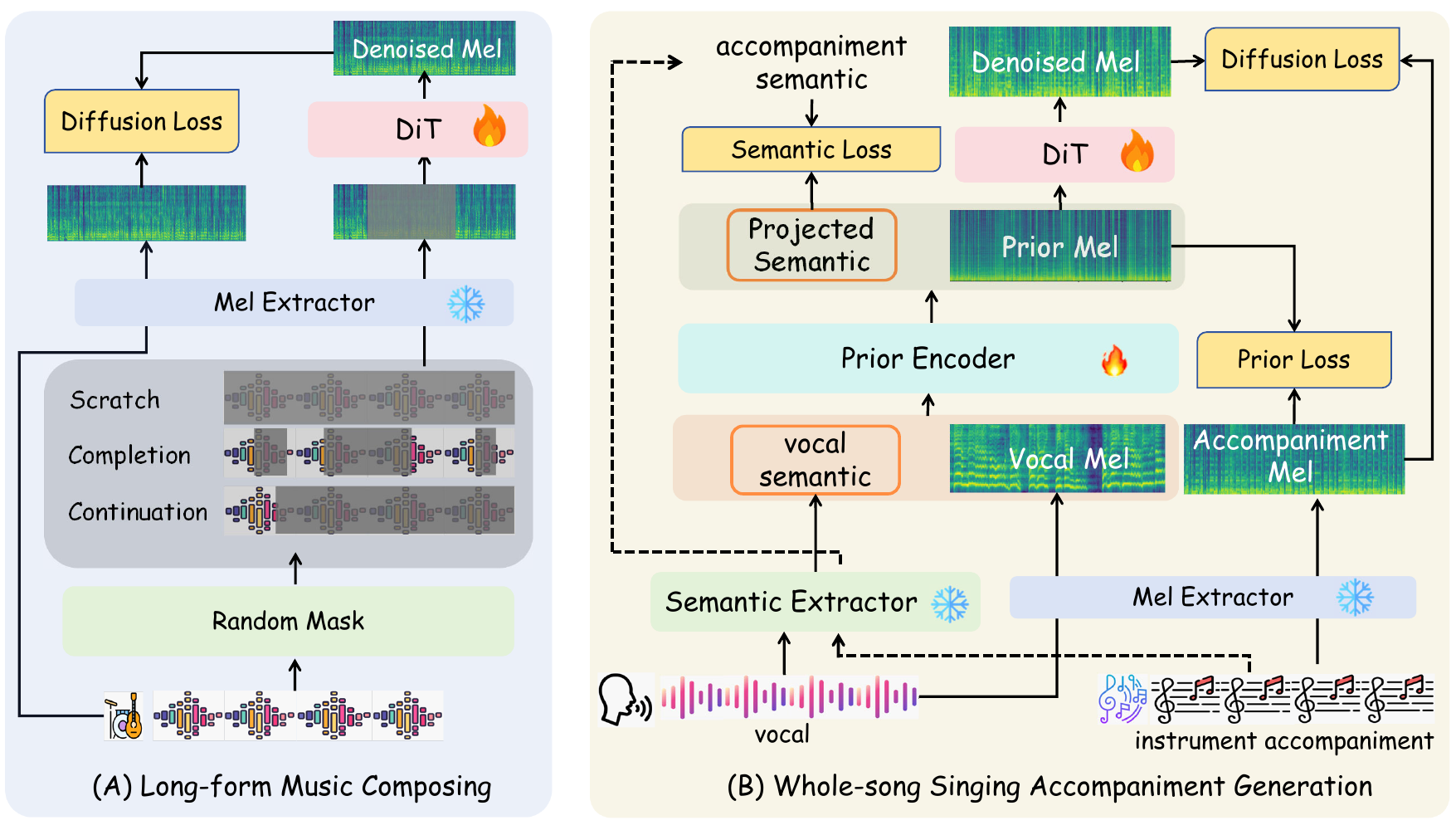}
\caption{Overview of the frameworks using SqueezeComposer for composing. (A) Long-form Music Composing: temporal speed-up enables efficient abstract-level generation using DiT with masking strategies (scratch, completion, continuation). (B) Whole-song Singing Accompaniment Generation: semantic-to-prior mapping followed by DiT refinement for harmonious accompaniment.}
\label{fig:task}
\end{figure*}

The training objective for the speed-up and restoration process consists of two components: 1) a prior loss between the CNN output and the ground truth Mel spectrogram, which guides the prior generation:
\begin{align}
\mathcal{L}_{\text{Prior}} = \mathbb{E}_{x} \left[ \left\| f_{\text{CNN}}(m_c, r) - m_0 \right\|_2^2 \right],    
\end{align}
where $m_c$ is the Mel spectrogram of the accelerated audio, $r$ is the scaling factor, $m_0$ is the ground truth Mel spectrogram, and $f_{\text{CNN}}$ is the CNN prior generator. 2) A refinement loss between the DiT output and the ground truth, which ensures high-quality restoration:
\begin{align}
\mathcal{L}_{\text{Refine}} = \mathbb{E}_{x, t} \left[ \left\| f_{\text{DiT}}(m^{(t)}, t, c) - m_0 \right\|_2^2 \right],    
\end{align}
where $m^{(t)}$ is the noisy version of the ground truth at timestep $t$, $f_{\text{DiT}}$ is the DiT model that predicts $m_0$ directly, and $c = f_{\text{CNN}}(m_c)$ is the CNN prior condition. This dual-loss strategy enables the model to learn both the  characteristics at the larger temporal-squeezed scale and the restoration quality effectively.


 
\section{Music Composition}
With the above designs, SqueezeComposer enables efficient long-form music generation, addressing key challenges in computational efficiency and musical structure modeling. To evaluate its effectiveness for music composition, we explore two aspects: temporal-wise generation for long-range structure and track-wise generation for multi-track coordination. For temporal-wise generation, the framework generates music spanning several or even over ten minutes, including continuation, completion, and generation from scratch. For track-wise generation, the framework generates instrumental tracks that align with a given vocal melody for a whole song. After generating in the temporal speed-up domain, SqueezeComposer employs temporal restoration (see Section \ref{restoring}) to recover the audio to its original speed, completing the full compress-and-expand pipeline. SqueezeComposer's key innovation lies in the hypothesis that AI models can understand and generate music at speeds far beyond human listening capabilities, where accelerated music corresponds to musical abstraction, naturally aligning with hierarchical generation principles.

\subsection{Long-form Music Composing}
For long-form music composition, temporally accelerated audio is first generated by backbone generation models and then restored to normal speed, enabling efficient generation while preserving structural information. For instance, when using the 8$\times$ speed-up, a model that could originally handle 2-minute audios due to memory and context modeling limitations can now produce 16-minute audios, achieving long-term generation with the simple trick of temporal squeezing.

Specifically, as shown in Fig.~\ref{fig:task} (A), three long-form music generation tasks are considered: continuation (predicting the subsequent audios), completion (bridging between audios or re-create certain intervals), and generation from scratch. A masking-based strategy is designed to achieve these tasks by masking the latter portion of the Mel spectrogram, intermediate sections, or the whole spectrogram, for the three tasks, respectively. The masked Mel spectrogram serves as the conditions for the DiT to predict the original spectrogram, optimized by the diffusion loss functions.


\subsection{Whole-song singing accompaniment generation}
The singing accompaniment generation creates an accompaniment track to complement an existing vocal track. Beyond low-level rhythmic alignment, achieving conceptual-level coherence is essential for high-quality accompaniment generation. We design a semantic-to-prior mapping approach that projects vocal features into a prior representation, which is then refined using a DiT to generate harmonious accompaniment.

As illustrated in Fig.~\ref{fig:task} (B), the generation pipeline consists of three main stages:
1) We extract vocal semantic features using a pretrained model (MERT \cite{li2023mert} or MuQ \cite{zhu2025muq}) and compute the vocal Mel spectrogram from the input audio. 2) A prior encoder takes both the vocal semantic features and the vocal Mel spectrogram as input to generate a prior Mel spectrogram, which serves as an initial estimate of the accompaniment. The vocal semantic features and vocal Mel spectrogram are fused through a bidirectional cross-attention mechanism \cite{Hiller2024PerceivingLS}. 3) The DiT refines this prior Mel spectrogram through a denoising process to produce the final accompaniment Mel spectrogram.
All processing is performed in the time-accelerated domain, enabling efficient whole-song generation (typically under 5 minutes) and allowing the model to learn song-level patterns, including intro, outro, and vocal-silence segments. The prior encoder uses 6 cross-attention layers, the DiT uses 8 layers, and the entire model is trained for 500k steps using 8 H800 GPUs.

The training process involves three loss functions to ensure high-quality generation. The semantic loss ensures semantic consistency by comparing the projected semantic representation with the target accompaniment semantic. The Prior Loss compares the generated Prior Mel with the target accompaniment Mel, guiding the Prior Encoder to produce accurate initial estimates, and is formulated as a mean squared error (MSE) loss. The Diffusion Loss trains the DiT to refine the Prior Mel into high-quality accompaniment that maintains harmony with the vocal input. This multi-stage approach enables effective cross-modal generation while preserving the temporal alignment and musical coherence between vocal and accompaniment tracks. 

\section{Experiments}
\subsection{Dataset}

\begin{figure}[!t]
\centering
\includegraphics[width=0.99\columnwidth]{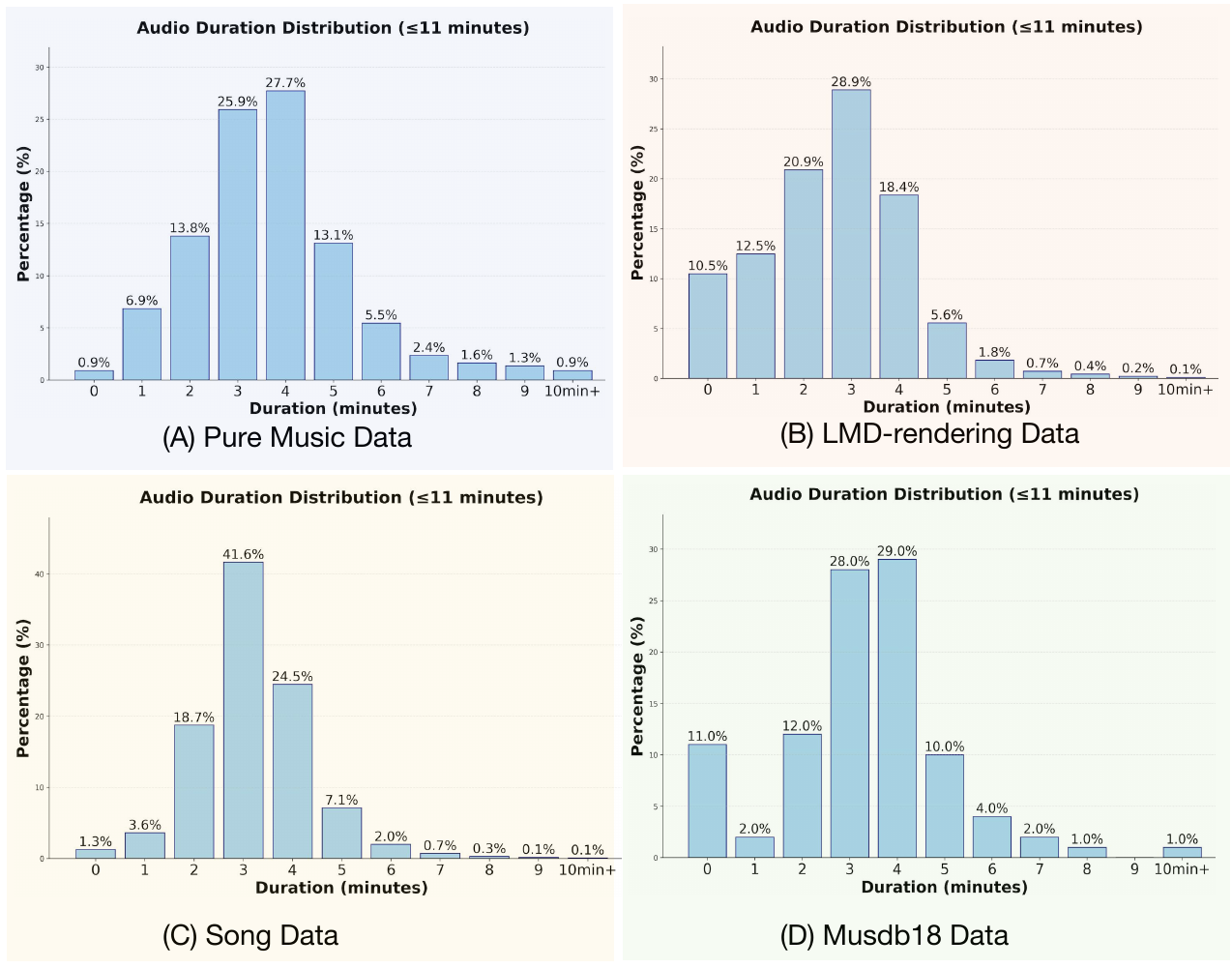}
\caption{Audio duration distribution across four music datasets. All datasets show concentration in the 3-10 minute range, demonstrating that the majority of music files fall within the long-form category, with peaks around 5-7 minutes depending on the dataset type.}
\label{fig:data}
\end{figure}

\begin{table*}[!t]
\caption{Comparison of temporal speed-up restoration quality against existing audio intermediate representations. All audio vocoders are off-the-shelf vocoders without any retraining.}
\label{tab:reconstruction1}
\centering
\begin{tabular}{lrrrrr}
\toprule
\textbf{Model} & \textbf{Mel$_{\text{dis}}$ ($\downarrow$)} & \textbf{STFT$_{\text{dis}}$ ($\downarrow$)} & \textbf{Sampling-rate} & \textbf{TPS} ($\downarrow$) & \textbf{F$_a$PS} ($\downarrow$) \\
\midrule
MuCodec \cite{xu2024mucodec} & 3.1204  & 0.4065 & 48,000 & 25 & 25 \\
Encodec \cite{defossez2023high} & 3.3829 & 0.3266 & 24,000 & 600 & 75 \\
DAC \cite{kumar2023high} & 2.1274 & 0.3682 & 24,000 & 450 & 50 \\
XCodec \cite{ye2025codec} & 2.4798 & 0.2263 & 16,000 & 400 & 50 \\
SNAC \cite{siuzdak2024snac} & 2.6088 & 0.2518 & 32,000 & 156 & - \\
Vocos \cite{siuzdak2023vocos} & 2.5959 & 0.3378 & 24,000 & - & 94.12 \\
\midrule
BigVGAN \cite{lee2023bigvgan} & 1.2115 & 0.5005 & 24,000 & -- & 93.75 \\
BigVGAN-Squeeze-4 & 2.9431 & 0.5214 & 24,000 & -- & 23.44 \\
BigVGAN-Squeeze-8 & 4.5814 & 0.8864 & 24,000 & -- & 11.72 \\
\bottomrule
\end{tabular}
\end{table*}

\begin{table*}[!t]
\caption{Comparison of temporal speed-up restoration quality with vocoder fine-tuning. $\star$ indicates that the vocoder is fine-tuned using Mel spectrograms restored from temporally squeezed inputs.}
\label{tab:reconstruction2}
\centering
\begin{tabular}{lrrrrr}
\toprule
\textbf{Model} & \textbf{Mel$_{\text{dis}}$ ($\downarrow$)} & \textbf{STFT$_{\text{dis}}$ ($\downarrow$)} & \textbf{Waveform$_{\text{dis}}$ ($\downarrow$)} & \textbf{F$_a$PS} ($\downarrow$) \\
\midrule
BigVGAN \cite{lee2023bigvgan} & 1.2115 & 0.5005 & 0.1575 & 93.75 \\
BigVGAN-Squeeze-4($\star$) & 1.8367 & 0.5414 & 0.1798 & 23.44 \\
BigVGAN-Squeeze-8($\star$) & 2.2512 & 0.6328 & 0.1822 & 11.72 \\
\bottomrule
\end{tabular}
\end{table*}

We use four different music datasets to demonstrate SqueezeComposer's effectiveness across various long-form music generation scenarios:
a) Pure Music Data: A curated dataset of approximately 7,000 pieces, including light music, classical guzheng and piano pieces, and classical compositions.
b) Lakh MIDI Dataset~\cite{raffel2016lakh}: Instrumental music rendered from MIDI to WAV format following~\cite{chen2024pyramidcodec}, containing approximately 160,000 pieces.
c) Song Data: A private dataset of approximately 400,000 songs, processed through source separation with demucs \cite{rouard2022hybrid} to extract vocals and accompaniment tracks.
d) MUSDB18 \cite{musdb18}: An open-source dataset that provides direct access to separated vocal and instrumental tracks.

Datasets a), b), and c) are split into training and testing sets with an 8:2 ratio. Datasets a) and b) are used for training long-form music generation tasks, including continuation, completion, and generation from scratch. Dataset c) is used for training whole-song singing accompaniment generation. Dataset d) is not used for training; its official test set is used to evaluate out-of-domain generation performance. The duration distribution is shown in Figure \ref{fig:data}. All datasets demonstrate concentration in the 3-10 minute range, confirming that the majority of music files fall within the long-form category, making them suitable for training and evaluating our temporal speed-up approach.

\subsection{Baseline and Metrics}
\textbf{Squeezing and Restoration.} For temporal speed-up and restoration evaluation, we compare against existing intermediate representations including Mel spectrograms with vocoders (Bigvgan\cite{lee2023bigvgan}, Vocos \cite{siuzdak2023vocos}) and codec representations (Encodec \cite{defossez2023high}, DAC \cite{kumar2023high}, Xcodec \cite{ye2025codec}, SNAC \cite{siuzdak2024snac}). We use Mel distance (L1 distance between Mel spectrograms with 2048-point FFT, 512-point hop size, and 80 Mel frequency bins), STFT distance (L1 distance between log magnitude spectrograms with the same FFT and hop size) and waveform distance (L1 distance between ground truth and reconstructed waveforms) to assess the reconstruction quality, along with TPS and audio frames per second (F$_a$PS) to measure the computational efficiency.

\textbf{Long-form Music Generation.} We compare SqueezeComposer against PyramidCodec \cite{chen2024pyramidcodec}, MusicGen \cite{copet2023simple}, audioLDM \cite{liu2024audioldm} and musicLDM \cite{chen2024musicldm} for long-form music generation tasks. The PyramidCodec is used for continuation and generation from scratch since it is used to train an AR model for the learned codecs, making the completion unfeasible.

\textbf{Whole-song Singing Accompaniment Generation.} We use SingSong~\cite{donahue2023singsong} and FastSAG~\cite{chen2024fastsag} as baselines for this task. We also use RandSong, which randomly selects accompaniments from a library of 20,000 pieces, to evaluate the importance of accompaniment harmony.

For both long-form music generation and whole-song SAG, we employ Fréchet Audio Distance (FAD) with~\cite{fadtk} and AudioBox-Aesthetics~\cite{tjandra2025aes} for evaluation. We use VGGish~\cite{hershey2017cnn} as the pretrained embedding model for FAD computation, resulting in the evaluation metric $\texttt{FAD}_{\texttt{VGGish}}$. AudioBox-Aesthetics introduces new tools for evaluating audio aesthetics across four dimensions: content enjoyment ($\texttt{CE}$), content usefulness ($\texttt{CU}$), production complexity ($\texttt{PC}$), and production quality ($\texttt{PQ}$). We also assess $\texttt{CE}$, $\texttt{CU}$, $\texttt{PC}$, and $\texttt{PQ}$ using this tool. For both tasks, we evaluate the generated duration, and for long-form music generation, we additionally measure real-time factor (RTF) to assess generation efficiency.

\subsection{Results}
\subsubsection{Squeezing and Restoration}

\begin{figure*}[!t]
\centering
\includegraphics[width=0.9\textwidth]{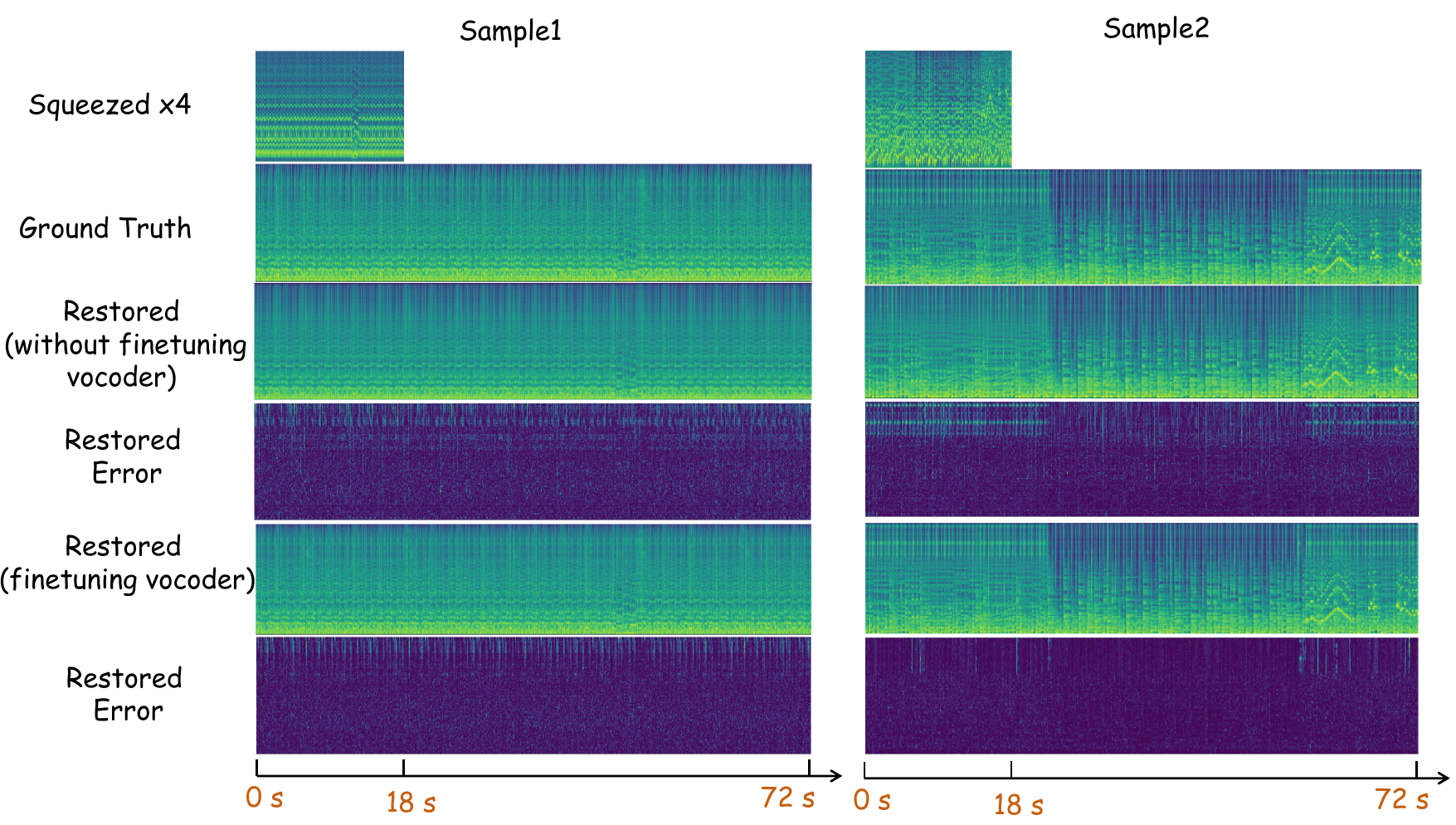}
\caption{Visualization of squeezing and ($\times$4) restoration results on two example samples. From top to bottom: the temporally squeezed Mel spectrogram, the original ground-truth Mel spectrogram, the restored Mel spectrogram using a pretrained vocoder, and the corresponding restoration error; the restored Mel spectrogram using a vocoder fine-tuned on restored Mel spectrogram, and its corresponding error. }
\label{fig:samples}
\end{figure*}

\begin{table*}[!t]
\caption{Evaluation Results for Continuation, Completion and from Scratch on rendered Lakh MIDI Dataset.}
\label{tab:continuation}
\centering
\resizebox{\textwidth}{!}{%
\begin{tabular}{lrrrrrrr}
\toprule
\textbf{Model} & \textbf{CE}($\uparrow$) & \textbf{CU}($\uparrow$) & \textbf{PC}($\uparrow$) & \textbf{PQ}($\uparrow$) & \textbf{FAD(vgg)}($\downarrow$) & \textbf{Duration}($\uparrow$) & \textbf{RTF}($\downarrow$) \\
\midrule
PyramidCodec \cite{chen2024pyramidcodec} (continuation) & 6.6442 & 7.9427 & 4.2218 & 8.1208 & 1.5201 & 180s & 10.490 \\
SqueezeComposer\_x4 (continuation) & 6.8499 & 7.8684 & 4.9940 & 7.8934 & 1.2321 & 240s & 0.078 \\
SqueezeComposer\_x4 (completion) & 6.7321 & 7.8267 & 4.8574 & 7.9340 & 1.3452 & 240s & 0.081 \\
\midrule
PyramidCodec \cite{chen2024pyramidcodec} (scratch) & 6.6816 & 7.8873 & 4.4297 & 8.0238 & - & 180s & 11.192 \\
SqueezeComposer\_x4 (scratch) & 6.6919 & 7.4632 & 4.7583 & 7.9916 & - & 240s & 0.079 \\
\bottomrule
\end{tabular}%
}
\end{table*}

\begin{table*}[!t]
\caption{Evaluation Results for Pure Instrumental Music Generation from Scratch.}
\label{tab:scratch}
\centering
\begin{tabular}{lrrrrrr}
\toprule
\textbf{Model} & \textbf{CE}($\uparrow$) & \textbf{CU}($\uparrow$) & \textbf{PC}($\uparrow$) & \textbf{PQ}($\uparrow$) & \textbf{Duration}($\uparrow$) & \textbf{RTF}($\downarrow$) \\
\midrule
MusicGen \cite{copet2023simple} & 4.4182 & 6.5688 & 4.3799 & 6.7619  & 15s & 6.518 \\
AudioLDM \cite{liu2024audioldm} & 6.4087 & 6.8974 & 4.1393 & 7.2572  & 60s & 0.005 \\
MusicLDM \cite{chen2024musicldm} & 4.8979 & 7.0832 & 4.8742 & 6.9630  & 15s & 0.011 \\
SqueezeComposer\_x4 & 6.4319 & 6.8824 & 5.5222 & 6.8856 & 240s & 0.074 \\
SqueezeComposer\_x8 & 6.0171 & 6.7128 & 5.0624 & 6.9417 & 600s & 0.037 \\
\bottomrule
\end{tabular}
\end{table*}

\begin{figure*}[!t]
\centering
\includegraphics[width=0.95\textwidth]{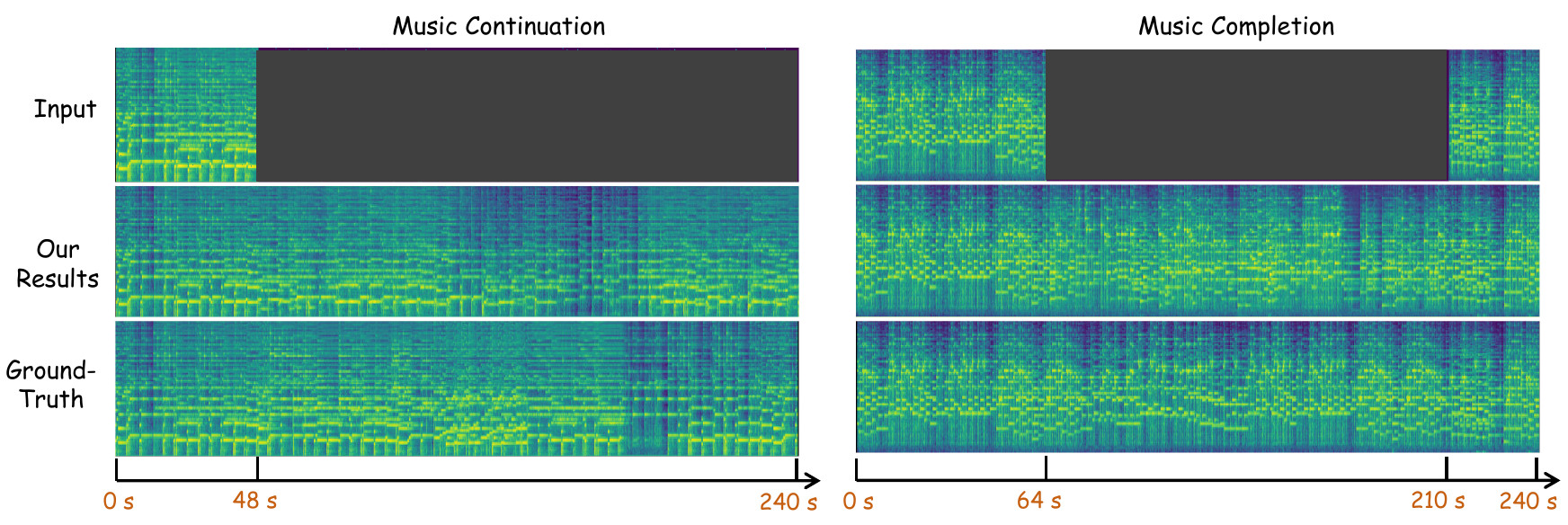}
\caption{Music continuation (left) and music completion (right) results using our temporal acceleration framework. All spectrograms are computed from audio that has been \textbf{accelerated by a factor of 4}, and the generation is performed entirely in this accelerated domain, enabling long-horizon modeling and efficient inference. \textbf{Top:} Input spectrograms with missing regions.  \textbf{Middle:} Our generated results, operating on the accelerated representation. \textbf{Bottom:} Ground-truth spectrograms for reference. }
\label{fig:continuation_completion}
\end{figure*}

The efficiency of intermediate representation extraction and decoding is crucial for scalable long-form music generation. Existing feature extraction and decoding models, such as Encodec, DAC, and BigvGAN vocoder, typically exhibit high TPS or F$_a$PS values, which can become computationally prohibitive for extended durations for music generation.

Table~\ref{tab:reconstruction1} reports the results using pretrained vocoders. Here, the BigVGAN-Squeeze models decode the restored Mel spectrogram from the proposed squeezing and restoration pipeline. It can be seen that even using the pretrained vocoders, for 4$\times$ acceleration, it could maintain perceptually consistent audio with only a slight degradation in objective metrics. Empirical study shows that the outputs for 8$\times$ is still quite reasonable even though objective results show an degradation. Moreover, the reduced F$_a$PS indicates that the strategy could substantially improve computational efficiency to enable scalable long-form generation. 

The vocoder fine-tuning is conducted with results shown in Table~\ref{tab:reconstruction2}. Specifically, the vocoder is retrained using the reconstructed Mel spectrograms obtained from the restoration model, to recover the original normal-speed waveform. We can notice that, the fine-tuning remarkably improves the reconstruction quality, effectively closing the gap introduced by acceleration, achieving performance on par with the original BigVGAN. They could also outperform other baselines in Table~\ref{tab:reconstruction1}. The visualization examples are also shown in Fig.~\ref{fig:samples} to further demonstrate that fine-tuning can reduce spectral distortion and high-frequency artifacts.

\subsubsection{Long-form Music Composing}

We evaluate the performance of SqueezeComposer on long-form music generation tasks, including continuation, completion, and generation from scratch, comparing it against established baselines.

From Table \ref{tab:continuation}, for continuation, completion and generation from scratch tasks on the rendered Lakh MIDI Dataset, our SqueezeComposer models demonstrate superior performance in terms of generation efficiency and duration compared to PyramidCodec. Our models achieve significantly lower Real-Time Factors (RTF), indicating much faster generation speed, and are capable of generating longer music segments. In terms of generation quality, our models achieve substantially lower FAD scores, indicating higher audio fidelity and similarity to real music. While PyramidCodec shows slightly higher scores in some aesthetic metrics, our models achieve competitive overall quality.

From Table \ref{tab:scratch}, for pure instrumental music generation from scratch, SqueezeComposer significantly outperforms MusicGen in terms of achievable duration and generation efficiency. MusicGen is limited to generating short 15-second clips with high RTF, while our SqueezeComposer models can generate music up to 10 minutes with very efficient RTF values with squeezing factor to be 8. This demonstrates SqueezeComposer's effectiveness in enabling truly long-form music generation, capable of producing extended compositions in approximately one minute. In terms of aesthetic metrics, our models achieve superior performance across all dimensions compared to MusicGen. These results highlight SqueezeComposer's ability to generate high-quality, coherent long-form music efficiently. AudioLDM and MuiscLDM, based on latent diffusion, achieves an extremely low RTF, reflecting its high generation efficiency; however, it is constrained to generating relatively short clips (up to 60 seconds). In contrast, SqueezeComposer not only supports substantially longer generation durations, but also achieves comparable or better performance on most metrics, indicating a more favorable trade-off between efficiency, generation length, and musical quality.

Furthermore, the spectrogram visualizations in Figure~\ref{fig:continuation_completion} provide intuitive evidence of the coherence achieved by SqueezeComposer. All spectrograms are shown in their temporally accelerated form (×4), meaning that both the input and the generated segments are compressed in time. Despite this acceleration, the generated regions in both the continuation and completion tasks exhibit smooth and consistent spectral evolution. For music continuation, the model extends the input with harmonically aligned and timbrally consistent patterns, without introducing abrupt spectral discontinuities. In the music completion setting—where a large middle segment is missing—the model successfully reconstructs a plausible continuation that closely matches the ground-truth trajectory in terms of harmonic density, rhythmic smoothness, and long-term spectral structure. These visual results further confirm that SqueezeComposer maintains strong temporal coherence and structural consistency throughout extended generation, even when operating entirely in the squeezed domain.

\begin{table*}[!t]
\caption{Evaluation Results for Singing Accompaniment Generation}
\label{tab:accompaniment}
\centering
\resizebox{\textwidth}{!}{%
\begin{tabular}{lllrrrrrr}
\toprule
\textbf{Dataset} & \textbf{Model} & \textbf{Config} & \textbf{CE}($\uparrow$) & \textbf{CU}($\uparrow$) & \textbf{PC}($\uparrow$) & \textbf{PQ}($\uparrow$) & \textbf{FAD}($\downarrow$) & \textbf{Duration}($\uparrow$) \\
\midrule
\multirow{3}{*}{MUSDB18} & FastSAG \cite{chen2024fastsag} & mixture & 6.5173 & 7.0275 & 6.2978 & 6.9921 & 1.2109 & 10s \\
& SingSong \cite{donahue2023singsong} & mixture & 6.1253 & 6.5430 & 5.9878 & 6.8324 & 0.9084 & 10s \\
& SqueezeComposer & mixture, x4, MERT & 5.6126 & 6.7193 & 6.8603 & 7.3931 & 4.1104 & 240s \\
\midrule
\multirow{8}{*}{In-domain} & RandSong & mixture & 5.0145 & 5.6639 & 3.0143 & 6.4618 & 21.6837 & 240s \\
& RandSong & instrument & 7.3543 & 7.8033 & 5.5412 & 7.8258 & 0.4331 & 240s \\
& FastSAG \cite{chen2024fastsag} & mixture & 6.2566 & 6.6753 & 6.7159 & 7.1718 & 1.7423 & 10s \\
& FastSAG \cite{chen2024fastsag} & instrument & 6.359 & 6.9085 & 6.2883 & 6.9243 & 3.0373 & 10s \\
& SqueezeComposer & mixture, x4, muQ & 7.0283 & 7.1914 & 6.5727 & 7.5715 & 1.1039 & 240s \\
& SqueezeComposer & instrument, x4, muQ & 6.9425 & 7.3406 & 5.8787 & 7.2001 & 1.8731 & 240s \\
& SqueezeComposer & mixture, x4, MERT & 7.0498 & 7.2146 & 6.4317 & 7.573 & 1.1899 & 240s \\
& SqueezeComposer & instrument, x4, MERT & 6.9761 & 7.3546 & 5.7592 & 7.2367 & 1.7073 & 240s \\
\bottomrule
\end{tabular}%
}
\end{table*}

\subsubsection{Whole-song Singing Accompaniment Generation}
We evaluate SqueezeComposer on whole-song singing accompaniment generation tasks, comparing it against established baselines including SingSong, FastSAG, and RandSong across both out-of-domain and in-domain scenarios.
Notably, when a method fails to generate a complete accompaniment for an entire song, segment-by-segment generation cannot guarantee rhythmic and stylistic consistency across the independently produced sections, as demonstrated on the project demo page.

For out-of-domain evaluation on MUSDB18, our SqueezeComposer model demonstrates the capability to generate significantly longer accompaniments compared to baseline models, enabling truly whole-song accompaniment generation. While our model achieves competitive performance in some AudioBox-Aesthetics metrics, it shows higher FAD scores compared to baselines, indicating room for improvement in audio fidelity.

For in-domain evaluation, our ablation study reveals several important insights. First, instrument-only tracks generally achieve better performance than mixture tracks across most metrics, suggesting that focusing on instrumental accompaniment generation leads to cleaner outputs. Second, different semantic encoders (MERT vs muQ) show comparable performance, indicating that the choice of semantic encoder has minimal impact on overall generation quality. All models consistently outperform RandSong in most metrics, demonstrating the effectiveness of our temporal speed-up approach for singing accompaniment generation.

The results demonstrate that SqueezeComposer successfully addresses the challenge of whole-song singing accompaniment generation, achieving significantly longer generation durations while maintaining reasonable quality across multiple evaluation dimensions.

\section{Conclusion}
In this work, we propose SqueezeComposer, a simple yet effective framework for long-form music generation that leverages temporal speed-up to address both computational challenges and musical structure complexity. Our key insight is that AI models can understand and generate time-accelerated audio, enabling efficient abstract-to-detail generation through hierarchical compression and expansion. By working in the time-accelerated domain, SqueezeComposer significantly reduces computational requirements while maintaining high-quality output and compatibility with existing vocoders.

\newpage
\bibliography{sample-base}

@String{ICLR = {Proc. Intl. Conf. Learning Representations (ICLR)}}

@String{ICML = {Proc. Intl. Conf. Machine Learning (ICML)}}

@String{NeurIPS = {Proc. Conf. Neural Information Processing Systems (NeurIPS)}}

@String{IEEE_TACM_TASLP = {IEEE/ACM Trans. Audio, Speech, Lang. Process.}}

@String{Interspeech = {Proc. Interspeech}}

@String{AAAI = {Proc. AAAI Conf. Artif. Intell. (AAAI)}}

@String{ACL = {Proc. Assoc. for Computational Linguistics (ACL)}}

@String{ISMIR              = {Proc. Intl. Soc. for Music Information Retrieval Conf. ({ISMIR})}}

@String{ICASSP = {Proc. IEEE Intl. Conf. Acoustics, Speech, Signal Process. (ICASSP)}}

@String{IJCAI = {Proc. Intl. Joint Conf. Artif. Intell. (IJCAI)}}

@String{ICCV = {Proc. IEEE/CVF Conf. Comput. Vis. Pattern Recogn (ICCV)}}

@String{EMNLP = {Proc. Conf. on Empirical Methods in Natural Language Processing (EMNLP)}}

@article{borsos2023audiolm,
  title={Audiolm: a language modeling approach to audio generation},
  author={Borsos, Zal{\'a}n and Marinier, Rapha{\"e}l and Vincent, Damien and Kharitonov, Eugene and Pietquin, Olivier and Sharifi, Matt and Roblek, Dominik and Teboul, Olivier and Grangier, David and Tagliasacchi, Marco and others},
  journal=IEEE_TACM_TASLP,
  year={2023}
}

@inproceedings{ye2025codec,
  title={Codec does matter: Exploring the semantic shortcoming of codec for audio language model},
  author={Ye, Zhen and Sun, Peiwen and Lei, Jiahe and Lin, Hongzhan and Tan, Xu and Dai, Zheqi and Kong, Qiuqiang and Chen, Jianyi and Pan, Jiahao and Liu, Qifeng and others},
  booktitle=AAAI,
  year={2025}
}

@article{siuzdak2024snac,
  title={Snac: Multi-scale neural audio codec},
  author={Siuzdak, Hubert and Gr{\"o}tschla, Florian and Lanzend{\"o}rfer, Luca A},
  journal={arXiv preprint arXiv:2410.14411},
  year={2024}
}

@inproceedings{van2017neural,
  title={Neural discrete representation learning},
  author={Van Den Oord, Aaron and Vinyals, Oriol and others},
  booktitle=NeurIPS,
  year={2017}
}

@inproceedings{goel2022s,
  title={It’s raw! audio generation with state-space models},
  author={Goel, Karan and Gu, Albert and Donahue, Chris and R{\'e}, Christopher},
  booktitle=ICML,
  year={2022}
}

@article{lee2025exploring,
  title={Exploring state-space-model based language model in music generation},
  author={Lee, Wei-Jaw and Hsieh, Fang-Chih and Chen, Xuanjun and Tsai, Fang-Duo and Yang, Yi-Hsuan},
  journal={arXiv preprint arXiv:2507.06674},
  year={2025}
}

@article{yuan2025diffusion,
  title={Diffusion-based symbolic music generation with structured state space models},
  author={Yuan, Shenghua and Tang, Xing and Chen, Jiatao and Xie, Tianming and Wang, Jing and Shi, Bing},
  journal={arXiv preprint arXiv:2507.20128},
  year={2025}
}

@article{erol2024audio,
  title={Audio mamba: Bidirectional state space model for audio representation learning},
  author={Erol, Mehmet Hamza and Senocak, Arda and Feng, Jiu and Chung, Joon Son},
  journal={IEEE Signal Processing Letters},
  year={2024},
  publisher={IEEE}
}

@inproceedings{gu2022efficiently,
  title={Efficiently Modeling Long Sequences with Structured State Spaces},
  author={Gu, Albert and Goel, Karan and R{\'e}, Christopher},
  booktitle=ICLR,
  year={2022}
}

@inproceedings{gu2020hippo,
  title={Hippo: Recurrent memory with optimal polynomial projections},
  author={Gu, Albert and Dao, Tri and Ermon, Stefano and Rudra, Atri and R{\'e}, Christopher},
  booktitle=NeurIPS,
  year={2020}
}

@inproceedings{gu2024mamba,
  title={Mamba: Linear-time sequence modeling with selective state spaces},
  author={Gu, Albert and Dao, Tri},
  booktitle={First conference on language modeling},
  year={2024}
}

@inproceedings{wu2025ts3,
  title={TS3-Codec: Transformer-Based Simple Streaming Single Codec},
  author={Wu, Haibin and Kanda, Naoyuki and Eskimez, Sefik Emre and Li, Jinyu},
  booktitle=Interspeech,
  year={2025}
}

@article{zhai2025one,
  title={One Quantizer is Enough: Toward a Lightweight Audio Codec},
  author={Zhai, Linwei and Ding, Han and Zhao, Cui and Wang, Ge and Zhi, Wang and Xi, Wei and others},
  journal={arXiv preprint arXiv:2504.04949},
  year={2025}
}

@article{li2025melcap,
  title={MelCap: A Unified Single-Codebook Neural Codec for High-Fidelity Audio Compression},
  author={Li, Jingyi and Zhao, Zhiyuan and Liu, Yunfei and Lin, Lijian and Zhu, Ye and Wu, Jiahao and Kong, Qiuqiang and Li, Yu},
  journal={arXiv preprint arXiv:2510.01903},
  year={2025}
}

@article{wang2025switchcodec,
  title={SwitchCodec: A High-Fidelity Nerual Audio Codec With Sparse Quantization},
  author={Wang, Jin and Jiang, Wenbin and Wang, Xiangbo and You, Yubo and Fang, Sheng},
  journal={arXiv preprint arXiv:2505.24437},
  year={2025}
}

@article{zhang2025mbcodec,
  title={MBCodec: Thorough disentangle for high-fidelity audio compression},
  author={Zhang, Ruonan and Hao, Xiaoyang and Han, Yichen and Cao, Junjie and Liu, Yue and Zhang, Kai},
  journal={arXiv preprint arXiv:2509.17006},
  year={2025}
}

@inproceedings{pepino2025encodecmae,
  title={EncodecMAE: Leveraging neural codecs for universal audio representation learning},
  author={Pepino, Leonardo and Riera, Pablo and Ferrer, Luciana},
  booktitle=Interspeech,
  year={2025}
}

@inproceedings{wu2023audiodec,
  title={Audiodec: An open-source streaming high-fidelity neural audio codec},
  author={Wu, Yi-Chiao and Gebru, Israel D and Markovi{\'c}, Dejan and Richard, Alexander},
  booktitle=ICASSP,
  year={2023}
}

@article{kumar2019melgan,
  title={Melgan: Generative adversarial networks for conditional waveform synthesis},
  author={Kumar, Kundan and Kumar, Rithesh and De Boissiere, Thibault and Gestin, Lucas and Teoh, Wei Zhen and Sotelo, Jose and De Brebisson, Alexandre and Bengio, Yoshua and Courville, Aaron C},
  journal=NeurIPS,
  year={2019}
}

@inproceedings{roberts2018hierarchical,
  title={A hierarchical latent vector model for learning long-term structure in music},
  author={Roberts, Adam and Engel, Jesse and Raffel, Colin and Hawthorne, Curtis and Eck, Douglas},
  booktitle=ICML,
  year={2018}
}

@inproceedings{dong2018musegan,
  title={Musegan: Multi-track sequential generative adversarial networks for symbolic music generation and accompaniment},
  author={Dong, Hao-Wen and Hsiao, Wen-Yi and Yang, Li-Chia and Yang, Yi-Hsuan},
  booktitle=AAAI,
  year={2018}
}

@inproceedings{huang2019music,
  title={Music transformer: Generating music with long-term structure},
  author={Huang, Cheng-Zhi Anna and Vaswani, Ashish and Uszkoreit, Jakob and Shazeer, Noam and Simon, Ian and Hawthorne, Curtis and Dai, Andrew M and Hoffman, Matthew D and Dinculescu, Monica and Eck, Douglas},
  booktitle=ICLR,
  year={2019}
}

@inproceedings{yu2022museformer,
  title={Museformer: Transformer with fine-and coarse-grained attention for music generation},
  author={Yu, Botao and Lu, Peiling and Wang, Rui and Hu, Wei and Tan, Xu and Ye, Wei and Zhang, Shikun and Qin, Tao and Liu, Tie-Yan},
  booktitle=NeurIPS,
  year={2022}
}

@inproceedings{hsiao2021compound,
  title={Compound word transformer: Learning to compose full-song music over dynamic directed hypergraphs},
  author={Hsiao, Wen-Yi and Liu, Jen-Yu and Yeh, Yin-Cheng and Yang, Yi-Hsuan},
  booktitle=AAAI,
  year={2021}
}

@inproceedings{mittal2021symbolic,
  title={Symbolic music generation with diffusion models},
  author={Mittal, Gautam and Engel, Jesse and Hawthorne, Curtis and Simon, Ian},
  booktitle=ISMIR,
  year={2021}
}

@inproceedings{evans2024fast,
  title={Fast timing-conditioned latent audio diffusion},
  author={Evans, Zach and Carr, CJ and Taylor, Josiah and Hawley, Scott H and Pons, Jordi},
  booktitle=ICML,
  year={2024}
}

@article{liu2024audioldm,
  title={Audioldm 2: Learning holistic audio generation with self-supervised pretraining},
  author={Liu, Haohe and Yuan, Yi and Liu, Xubo and Mei, Xinhao and Kong, Qiuqiang and Tian, Qiao and Wang, Yuping and Wang, Wenwu and Wang, Yuxuan and Plumbley, Mark D},
  journal=IEEE_TACM_TASLP,
  volume={32},
  pages={2871--2883},
  year={2024},
  publisher={IEEE}
}

@article{xu2024multi,
  title={Multi-source music generation with latent diffusion},
  author={Xu, Zhongweiyang and Dutta, Debottam and Wei, Yu-Lin and Choudhury, Romit Roy},
  journal={arXiv preprint arXiv:2409.06190},
  year={2024}
}

@inproceedings{yang2024uniaudio,
  title={Uniaudio: Towards universal audio generation with large language models},
  author={Yang, Dongchao and Tian, Jinchuan and Tan, Xu and Huang, Rongjie and Liu, Songxiang and Guo, Haohan and Chang, Xuankai and Shi, Jiatong and Bian, Jiang and Zhao, Zhou and others},
  booktitle=ICML,
  year={2024}
}

@inproceedings{melechovsky2024mustango,
  author    = {Jan Melechovský and Zixun Guo and Deepanway Ghosal and Navonil Majumder and Dorien Herremans and Soujanya Poria},
  title     = {Mustango: Toward Controllable Text-to-Music Generation},
  booktitle = {Proc. NAACL-HLT},
  year      = {2024},
  pages     = {8293--8316}
}

@inproceedings{chen2024musicldm,
  title={Musicldm: Enhancing novelty in text-to-music generation using beat-synchronous mixup strategies},
  author={Chen, Ke and Wu, Yusong and Liu, Haohe and Nezhurina, Marianna and Berg-Kirkpatrick, Taylor and Dubnov, Shlomo},
  booktitle=ICASSP,
  year={2024}
}

@inproceedings{evans2024long,
  title={Long-form music generation with latent diffusion},
  author={Evans, Zach and Parker, Julian D and Carr, CJ and Zukowski, Zack and Taylor, Josiah and Pons, Jordi},
  booktitle=ISMIR,
  year={2024}
}

@article{zhu2023ernie,
  title={Ernie-music: Text-to-waveform music generation with diffusion models},
  author={Zhu, Pengfei and Pang, Chao and Chai, Yekun and Li, Lei and Wang, Shuohuan and Sun, Yu and Tian, Hao and Wu, Hua},
  journal={arXiv preprint arXiv:2302.04456},
  year={2023}
}

@article{tjandra2025aes,
    title={Meta Audiobox Aesthetics: Unified Automatic Quality Assessment for Speech, Music, and Sound},
    author={Andros Tjandra and Yi-Chiao Wu and Baishan Guo and John Hoffman and Brian Ellis and Apoorv Vyas and Bowen Shi and Sanyuan Chen and Matt Le and Nick Zacharov and Carleigh Wood and Ann Lee and Wei-Ning Hsu},
    year={2025},
    journal = {arXiv preprint arXiv:2502.05139},
}

@inproceedings{hershey2017cnn,
  title={CNN architectures for large-scale audio classification},
  author={Hershey, Shawn and Chaudhuri, Sourish and Ellis, Daniel PW and Gemmeke, Jort F and Jansen, Aren and Moore, R Channing and Plakal, Manoj and Platt, Devin and Saurous, Rif A and Seybold, Bryan and others},
  booktitle=ICASSP,
  year={2017}
}

@article{Hiller2024PerceivingLS,
    title   = {Perceiving Longer Sequences With Bi-Directional Cross-Attention Transformers},
    author  = {Markus Hiller and Krista A. Ehinger and Tom Drummond},
    journal = {arXiv preprint arXiv:2402.12138},
    year    = {2024},
}

@inproceedings{fadtk,
  title = {Adapting Frechet Audio Distance for Generative Music Evaluation},
  author = {Gui, Azalea and Gamper, Hannes and Braun, Sebastian and Emmanouilidou, Dimitra},
  booktitle = ICASSP,
  year = {2024}
}

@inproceedings{rouard2022hybrid,
  title={Hybrid Transformers for Music Source Separation},
  author={Rouard, Simon and Massa, Francisco and D{\'e}fossez, Alexandre},
  booktitle=ICASSP,
  year={2023}
}

@misc{musdb18,
  author       = {Rafii, Zafar and
                  Liutkus, Antoine and
                  Fabian-Robert St{\"o}ter and
                  Mimilakis, Stylianos Ioannis and
                  Bittner, Rachel},
  title        = {The {MUSDB18} corpus for music separation},
  month        = dec,
  year         = 2017,
  doi          = {10.5281/zenodo.1117372},
  url          = {https://doi.org/10.5281/zenodo.1117372}
}

@misc{raffel2016lakh,
  title={The Lakh MIDI Dataset v0.1},
  author={Raffel, Colin},
  year={2016},
  howpublished={\url{https://colinraffel.com/projects/lmd/}},
  note={Accessed: [date]}
}

@inproceedings{li2023mert,
  title={{MERT}: Acoustic music understanding model with large-scale self-supervised training},
  author={Li, Yizhi and Yuan, Ruibin and Zhang, Ge and Ma, Yinghao and Chen, Xingran and Yin, Hanzhi and Xiao, Chenghao and Lin, Chenghua and Ragni, Anton and Benetos, Emmanouil and others},
  booktitle=ICLR,
  year={2024}
}

@article{zhu2025muq,
  title={{MuQ}: Self-supervised music representation learning with mel residual vector quantization},
  author={Zhu, Haina and Zhou, Yizhi and Chen, Hangting and Yu, Jianwei and Ma, Ziyang and Gu, Rongzhi and Luo, Yi and Tan, Wei and Chen, Xie},
  journal={arXiv preprint arXiv:2501.01108},
  year={2025}
}

@inproceedings{song2021score,
  title={Score-based generative modeling through stochastic differential equations},
  author={Song, Yang and Sohl-Dickstein, Jascha and Kingma, Diederik P and Kumar, Abhishek and Ermon, Stefano and Poole, Ben},
  booktitle=ICLR,
  year={2021}
}

@inproceedings{peebles2023scalable,
  title={Scalable diffusion models with transformers},
  author={Peebles, William and Xie, Saining},
  booktitle=ICCV,
  year={2023}
}

@inproceedings{chen2024fastsag,
  title={FastSAG: towards fast non-autoregressive singing accompaniment generation},
  author={Chen, Jianyi and Xue, Wei and Tan, Xu and Ye, Zhen and Liu, Qifeng and Guo, Yike},
  booktitle=IJCAI,
  year={2024}
}

@inproceedings{siuzdak2023vocos,
  title={Vocos: Closing the gap between time-domain and fourier-based neural vocoders for high-quality audio synthesis},
  author={Siuzdak, Hubert},
  booktitle=ICLR,
  year={2024}
}

@inproceedings{chen2024deep,
  title={Deep compression autoencoder for efficient high-resolution diffusion models},
  author={Chen, Junyu and Cai, Han and Chen, Junsong and Xie, Enze and Yang, Shang and Tang, Haotian and Li, Muyang and Lu, Yao and Han, Song},
  booktitle=ICLR,
  year={2025}
}

@article{gong2025ace,
  title={Ace-step: A step towards music generation foundation model},
  author={Gong, Junmin and Zhao, Sean and Wang, Sen and Xu, Shengyuan and Guo, Joe},
  journal={arXiv preprint arXiv:2506.00045},
  year={2025}
}

@inproceedings{jiang2025unicodec,
  title={{UniCodec}: Unified Audio Codec with Single Domain-Adaptive Codebook},
  author={Jiang, Yidi and Chen, Qian and Ji, Shengpeng and Xi, Yu and Wang, Wen and Zhang, Chong and Yue, Xianghu and Zhang, ShiLiang and Li, Haizhou},
  booktitle=ACL,
  year={2025}
}

@article{xu2024mucodec,
  title={Mucodec: Ultra low-bitrate music codec},
  author={Xu, Yaoxun and Chen, Hangting and Yu, Jianwei and Tan, Wei and Gu, Rongzhi and Lei, Shun and Lin, Zhiwei and Wu, Zhiyong},
  journal={arXiv preprint arXiv:2409.13216},
  year={2024}
}

@article{defossez2023high,
  author = {Alexandre D{\'e}fossez and Jade Copet and Gabriel Synnaeve and Yossi Adi},
  title = {High Fidelity Neural Audio Compression},
  journal = {Trans. Mach. Learn. Res.},
  year = {2023},
  volume = {2023}
}

@inproceedings{kumar2023high,
  title={High-fidelity audio compression with improved rvqgan},
  author={Kumar, Rithesh and Seetharaman, Prem and Luebs, Alejandro and Kumar, Ishaan and Kumar, Kundan},
  booktitle=NeurIPS,
  year={2023}
}

@inproceedings{kong2020hifi,
  title={Hifi-gan: Generative adversarial networks for efficient and high fidelity speech synthesis},
  author={Kong, Jungil and Kim, Jaehyeon and Bae, Jaekyoung},
  booktitle=NeurIPS,
  year={2020}
}

@inproceedings{lee2023bigvgan,
  title={{BigVGAN}: A Universal Neural Vocoder with Large-Scale Training},
  author={Lee, Sang-gil and Ping, Wei and Ginsburg, Boris and Catanzaro, Bryan and Yoon, Sungroh},
  booktitle=ICLR,
  year={2023}
}

@inproceedings{lam2023efficient,
  title={Efficient neural music generation},
  author={Lam, Max WY and Tian, Qiao and Li, Tang and Yin, Zongyu and Feng, Siyuan and Tu, Ming and Ji, Yuliang and Xia, Rui and Ma, Mingbo and Song, Xuchen and others},
  booktitle=NeurIPS,
  year={2023}
}

@inproceedings{chen2024pyramidcodec,
  title={Pyramidcodec: Hierarchical codec for long-form music generation in audio domain},
  author={Chen, Jianyi and Dai, Zheqi and Ye, Zhen and Tan, Xu and Liu, Qifeng and Guo, Yike and Xue, Wei},
  booktitle=EMNLP,
  year={2024}
}

@article{donahue2023singsong,
  title={Singsong: Generating musical accompaniments from singing},
  author={Donahue, Chris and Caillon, Antoine and Roberts, Adam and Manilow, Ethan and Esling, Philippe and Agostinelli, Andrea and Verzetti, Mauro and Simon, Ian and Pietquin, Olivier and Zeghidour, Neil and others},
  journal={arXiv preprint arXiv:2301.12662},
  year={2023}
}

@article{agostinelli2023musiclm,
  title={Musiclm: Generating music from text},
  author={Agostinelli, Andrea and Denk, Timo I and Borsos, Zal{\'a}n and Engel, Jesse and Verzetti, Mauro and Caillon, Antoine and Huang, Qingqing and Jansen, Aren and Roberts, Adam and Tagliasacchi, Marco and others},
  journal={arXiv preprint arXiv:2301.11325},
  year={2023}
}

@inproceedings{copet2023simple,
  title={Simple and controllable music generation},
  author={Copet, Jade and Kreuk, Felix and Gat, Itai and Remez, Tal and Kant, David and Synnaeve, Gabriel and Adi, Yossi and D{\'e}fossez, Alexandre},
  booktitle=NeurIPS,
  year={2023}
}

@inproceedings{schneider2024mousai,
  title={Mo{\^u}sai: Efficient text-to-music diffusion models},
  author={Schneider, Flavio and Kamal, Ojasv and Jin, Zhijing and Sch{\"o}lkopf, Bernhard},
  booktitle=ACL,
  year={2024}
}

@article{huang2023noise2music,
  title={Noise2music: Text-conditioned music generation with diffusion models},
  author={Huang, Qingqing and Park, Daniel S and Wang, Tao and Denk, Timo I and Ly, Andy and Chen, Nanxin and Zhang, Zhengdong and Zhang, Zhishuai and Yu, Jiahui and Frank, Christian and others},
  journal={arXiv preprint arXiv:2302.03917},
  year={2023}
}

@article{dhariwal2020jukebox,
  title={Jukebox: A generative model for music},
  author={Dhariwal, Prafulla and Jun, Heewoo and Payne, Christine and Kim, Jong Wook and Radford, Alec and Sutskever, Ilya},
  journal={arXiv preprint arXiv:2005.00341},
  year={2020}
}

@article{zeghidour2022soundstream,
  author = {Neil Zeghidour and Alejandro Luebs and Ahmed Omran and Jan Skoglund and Marco Tagliasacchi},
  title = {SoundStream: An End-to-End Neural Audio Codec},
  journal =  IEEE_TACM_TASLP,
  volume = {30},
  pages = {495--507},
  year = {2022}
}
\bibliographystyle{ACM-Reference-Format}







\end{document}